\documentclass[10pt,letterpaper]{article}
\usepackage{opex3}
\usepackage{amsmath,SIunits}
\usepackage{dsfont}

\begin{document}

\title{Two-photon interference with continuous-wave multi-mode coherent light}

\author{Yong-Su Kim$^{1,2,*}$, Oliver Slattery$^1$, Paulina S. Kuo$^1$, and Xiao Tang$^{1,\dag}$}
\address{$^1$Information Technology Laboratory, National Institute of Standards and Technology, \\100 Bureau Dr., Gaithersburg, MD 20899, USA\\
$^2$Center for Nano $\&$ Quantum Information, Korea Institute of Science and Technology (KIST), Seoul 136-791, Korea}
\email{$^*$yong-su.kim@kist.re.kr,~$^\dag$xiao.tang@nist.gov}

\date{\today}

\begin{abstract}
We report two-photon interference with continuous-wave multi-mode coherent light. We show that the two-photon interference, in terms of the detection time difference, reveals two-photon beating fringes with the visibility $V=0.5$. While scanning the optical delay of the interferometer, Hong-Ou-Mandel dips or peaks are measured depending on the chosen detection time difference. The HOM dips/peaks are repeated when the optical delay and the first-order coherence revival period of the multi-mode coherent light are the same.
\end{abstract}





\section{Introduction}

Interference is one of the most fascinating topics in optical physics. Since its first implementation by Young \cite{young}, many interesting phenomena have been studied \cite{hecht}. Both in classical and quantum physics, the key notion to understand interference is the superposition principle. In classical physics, light is considered as electromagnetic waves and the superposition of these waves explains the interference effects. On the other hand, when a photon, a light quantum, is considered in quantum physics, the superposition of probability amplitudes describes the interference. Despite the different analogies, both classical and quantum descriptions typically give the same results for first-order or single-photon interference. For example, a Young's double-slit interference experiment performed with either coherent light or single-photon states show the same interference fringes.

These results, however, become different when second-order or two-photon interference is explored \cite{mandel99}. Let us consider a typical two-photon interference scenario, the Hong-Ou-Mandel (HOM) interference \cite{hom}. When two identical optical pulses are combined at a beamsplitter (BS), the photons (or optical pulses in classical physics) are inclined to be detect at the same output of the BS, thus resulting in suppression of coincidences between the two outputs of the BS. This phenomenon is referred as a HOM dip, and the visibility of the HOM dip is defined as the relative depth of the dip compared to the non-interfering terms. Using single-photon states, the coincidences can be completely suppressed, so the visibility can reach $V=1$. On the other hand, classical electromagnetic waves superposition theory provides that a HOM dip with only $V\le0.5$ \cite{rarity05}. Thus, the visibility of $V=0.5$ in HOM interference is usually considered as the border between classical and quantum physics.

Because of the fundamental interest in the {\it quantum} nature of light, there has been a lot of research on two-photon quantum interference. These include two-photon coherence \cite{jha_thesis}, quantum beating \cite{ou88, legero03, legero04, legero06}, induced interference\cite{zou91}, and so on. The rapidly developing quantum information science has also boosted research on two-photon quantum interference \cite{kwon10, lim11, kim12}. Note that two-photon quantum interference is essential for many quantum information protocols including linear optics quantum computation \cite{klm, kok07}.

Recent research shows that the two-photon interference with classical light can sometimes imitate quantum interference and thus it can be useful for quantum information science. For example, ghost imaging or ghost interference, which was considered as a result of two-photon quantum interference, can be implemented with classical light sources \cite{wang04, cheng04, xiong05, zhai05}. Since the implementation of classical light sources is much easier than that of quantum light sources, these results show the practical benefits of two-photon classical interference for quantum information science. Thus, the study of two-photon classical interference is not only important for a better understanding of the nature of interference but also for applications in quantum information science.

In this paper, we theoretically and experimentally study two-photon interference with continuous-wave (CW) multi-mode coherent light. The remainder of the paper is organized as follows: we first introduce our scenario of two-photon interference with CW coherent light and provide a qualitative discussion. Then, we provide a quantitative theoretical analysis based on the superposition of electromagnetic waves. The theory is verified by experimental demonstration and results that follow. Finally, we will summarize our work and conclude.


\section{Two-photon interference with CW coherent light}

\begin{figure}[b]
\centering
\includegraphics[width=5.3in]{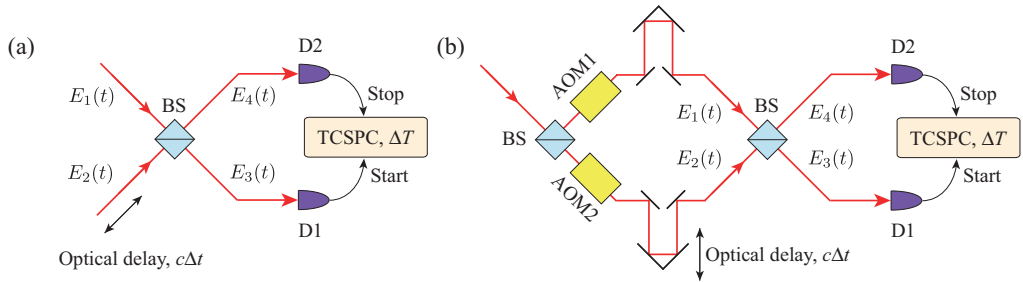}
\caption{(a) The schematic of two-photon interference with CW coherent light. (b) Experimental setup of two-photon interference with CW coherent light with a single laser. The degree of first-order coherence $|\gamma(\Delta t)|$ can be maintained to be nonzero while the phases between two inputs are randomized by two independent acousto-optic modulators, AOM1 and AOM2.}
\label{setup}
\end{figure}

Fig.~\ref{setup} (a) shows our scenario of two-photon interference with CW coherent light. The multi-mode property will be considered in the next section. Two attenuated CW coherent light beams interfere at a BS and are measured at the outputs of the BS. The two-photon measurement is accomplished by two single-photon detectors, D1and D2, and a time-correlated single-photon counter (TCSPC) which registers the detection time difference between two inputs; start and stop. For the interferometric measurement, we introduce two variables: $\Delta t$ and $\Delta T$ which denote the optical delay and the detection time difference at the TCSPC, respectively. Note that the optical delay $\Delta t$ can be changed by varying the optical length of an input.

Let us first consider the interference in terms of $\Delta t$. This two-photon interference is usually configured with optical pulses and the coincidences at D1 and D2 yield a HOM interference. Since the timing is well defined for optical pulses, $\Delta t$ would introduce an optical delay with respect to the reference optical pulse, $E_1(t)$ in Fig.~\ref{setup} (a). Thus, by varying $\Delta t$, one can expect to see two-photon interference with either single-photon states or coherent pulses \cite{hom, kim13}. Our interest in this paper is, however, CW coherent light. For CW light, the timing is not well defined, and therefore the meaning of the optical delay is ambiguous. Without varying the optical delay, it is hard to envision the interference phenomenon. For arbitrary CW coherent light inputs, the varying of $\Delta t$ does not change the arbitrary nature of the inputs and therefore it would not introduce interference. However, in contrast to intuition, we will show that the two-photon interference can be observed if a certain condition is satisfied.

The intuitive understanding of the two-photon classical interference in terms of $\Delta T$ is difficult. However, similar experiments with single-photon states can provide a hint. Legero, {\it et al} investigated the two-photon interference in terms of $\Delta T$ with single-photon states and observed two-photon quantum beating when the spectral frequencies of single-photon states are different \cite{legero03, legero04, legero06}. The beating fringes show a sinusoidal oscillation with the visibility of $V=1$. Based on this result, we can expect sinusoidal beat fringes for the two-photon classical interference with respect to $\Delta T$. As the visibility of two-photon classical interference is limited by $V=0.5$, we also can presume the visibility of the fringes would be $V\le0.5$. We will see the two-photon beating can indeed be observed with CW coherent light.


\section{Theoretical analysis with the superposition of electromagnetic waves}

Since we are dealing with classical light, we can describe two-photon interference phenomena by the superposition of electromagnetic waves. To this end, let us consider the electric fields $E_i(t)$ at the inputs ($i=1,2$) and outputs ($i=3,4$) of the BS as depicted in Fig.~\ref{setup} (a). Assuming that the amplitudes of $E_1(t)$ and $E_2(t)$ are the same, $|E_0|$, the two inputs of electric field are represented by
\begin{equation}
E_j(t)=|E_0|e^{i(\omega_j t+\phi_j)},
\label{inputs}
\end{equation}
where $j=1,2$, $\omega_j$ and $\phi_j$ are the angular frequencies and phases of $E_j$. The unitary transformation of a BS gives the output electric fields as 
\begin{eqnarray}
E_3(t)=\frac{1}{\sqrt{2}}[E_1(t-t_1)+iE_2(t-t_2)],\nonumber\\
E_4(t)=\frac{1}{\sqrt{2}}[E_2(t-t_2)+iE_1(t-t_1)],\label{BS_relation}
\end{eqnarray}
where $t_j$ is the time of flight from $E_j(t)$ to $E_3(t)$ and/or $E_4(t)$. Note that the relative time difference between $t_1$ and $t_2$ can be defined as the relative optical delay $\Delta t=t_2-t_1$. 

Although we represented the equations with the assumption of single-mode inputs, i.e., single frequencies $\omega_j$ for each input, there is typically non-zero spectral bandwidth. The non-zero spectral bandwidth produces a finite coherence length, which is determined by the degree of first-order coherence $0\le|\gamma(\Delta t)|\le1$ between $E_1(t)$ and $E_2(t+\Delta t)$, where $|\gamma(\Delta t)|=0$ for incoherent inputs while $|\gamma(\Delta t)|=1$ for completely coherent inputs. With the degree of first-order coherence $|\gamma(\Delta t)|$ and Eqs.~(\ref{inputs}) and (\ref{BS_relation}), the output intensities at D1 and D2 are given as
\begin{eqnarray}
I_3(t)=\langle E_3^*(t)E_3(t)\rangle=|E_0|^2\{1+|\gamma(\Delta t)|\langle\sin[\omega_2 t_2-\omega_1 t_1-\Delta\omega t-\Delta\phi]\rangle\},\nonumber\\
I_4(t)=\langle E_4^*(t)E_4(t)\rangle=|E_0|^2\{1-|\gamma(\Delta t)|\langle\sin[\omega_2 t_2-\omega_1 t_1-\Delta\omega t-\Delta\phi]\rangle\},
\label{I_34}
\end{eqnarray}
where $\Delta\omega$ and $\Delta\phi$ refer the frequency and phase difference between two inputs, i.e., $\Delta\omega=\omega_2-\omega_1$ and $\Delta\phi=\phi_2-\phi_1$. Here, $\langle x\rangle$ represents the average of $x$ over many events. When $E_1$ and $E_2$ have the same frequencies, $\omega_1=\omega_2=\omega_0$, the intensities $I_3$ and $I_4$ can be represented as
\begin{eqnarray}
I_3(t)&=&|E_0|^2\{1+|\gamma(\Delta t)|\langle\sin[\omega_0\Delta t-\Delta\phi]\rangle\},\nonumber\\
I_4(t)&=&|E_0|^2\{1-|\gamma(\Delta t)|\langle\sin[\omega_0\Delta t-\Delta\phi]\rangle\}.
\label{I_34_simple}
\end{eqnarray}
When the two inputs are coherent such that $\Delta\phi$ is a constant, Eq.~(\ref{I_34_simple}) shows sinusoidal oscillations with respect to $\Delta t$ with an envelope defined by $|\gamma(\Delta t)|$. This is single-photon interference. For incoherent inputs in which $\Delta\phi$ randomly varies, the sinusoidal oscillations will be washed out since $\langle \sin[\Delta\phi]\rangle=0$.

When the inputs are very weak, the coincidences between D1 and D2 correspond to the correlation measurement between $I_3(t)$ and $I_4(t)$. For a general description, let us consider that the detection times $T_1$ at D1 and $T_2$ at D2 are arbitrary. The detection time difference $\Delta T$ is related to these two detection times, $\Delta T=T_2-T_1$.

The intensity correlation between two detectors can be calculated as
\begin{equation}
\begin{aligned}
\langle I_{3}(T_1)I_4(T_2)\rangle=|E_0|^4&\{1
+|\gamma(\Delta t)|\langle\sin[\mathcal{A}-\Delta\omega T_1]\rangle
-|\gamma(\Delta t)|\langle\sin[\mathcal{A}-\Delta\omega T_2]\rangle\\
&-|\gamma(\Delta t)|^2\langle\sin[\mathcal{A}-\Delta\omega T_1]\sin[\mathcal{A}-\Delta\omega T_2\rangle\}.
\end{aligned}
\label{I_3I_4}
\end{equation}
where $\mathcal{A}=\omega_2t_2-\omega_1t_1-\Delta\phi$. For incoherent inputs, the second and third terms of Eq.~(\ref{I_3I_4}) vanish since $\langle\sin\mathcal{A}\rangle=0$. Note that the last term of Eq.~(\ref{I_3I_4}) does not disappear as it has a square of $\sin\mathcal{A}$ term. Dropping the constant $|E_0|^4$, Eq.~(\ref{I_3I_4}) can be simplified as
\begin{equation}
\langle I_{3}(T_1)I_4(T_2)\rangle\sim1-\frac{1}{2}|\gamma(\Delta t)|^2\langle\cos[\Delta\omega \Delta T]\rangle.
\label{I_3I_4_simple}
\end{equation}

Let us first determine the influence of Eq.~(\ref{I_3I_4_simple}) on the degree of first-order coherence, $|\gamma(\Delta t)|$. If two inputs of CW coherent light are completely independent and incoherent to each other from the beginning, $|\gamma(\Delta t)|=0$ at all times. In this case, no interference can be attained since the second term of Eq.~(\ref{I_3I_4_simple}) disappears. If, however, the inputs somehow have a non-zero first-order coherence, $|\gamma(\Delta t)|\neq0$, and have randomized phases, we can see interference. This condition can be achieved if, for instance, the two inputs originated from a single laser and their phases are randomized after they are separated. Fig.~\ref{setup} (b) shows a typical way to implement this condition. Note that two independent acousto-optic modulators, AOM1 and AOM2, disturb the phase coherence between two inputs, $E_1(t)$ and $E_2(t)$, and thus $\langle\sin[\Delta\phi]\rangle=0$.

When the frequencies of $E_1(t)$ and $E_2(t)$ are the same, $\Delta\omega=0$, the cosine term in Eq.~(\ref{I_3I_4_simple}) goes to 1 for any $\Delta T$. Assuming $|\gamma(\Delta t)|$ has a Gaussian distribution function, one can expect to observe a HOM dip with a visibility of $V=0.5$ while scanning $\Delta t$. It is remarkable that the HOM dip originated from the first-order coherence between the two inputs, even though the single-photon interference is erased by the randomized phases. Note that the HOM dip can be measured for any $\Delta T$, even when $\Delta T$ is much larger than the coherence time $t_c$ of the light source. This result is somewhat counter-intuitive since the coincidences for $t_c\ll\Delta T$ originate from photons that are temporally separated at the BS. As these photons did not have temporal overlap at the BS, one can naively think the electric fields do not interfere, thus they should not show interference. Eq.~(\ref{I_3I_4_simple}) shows that this intuition is incorrect and the electric fields do interfere even if they do not have temporal overlap. A similar discussion with coherent optical pulses can be found in Ref. \cite{kim13}.

\begin{figure}[t]
\centering
\includegraphics[width=3.2in]{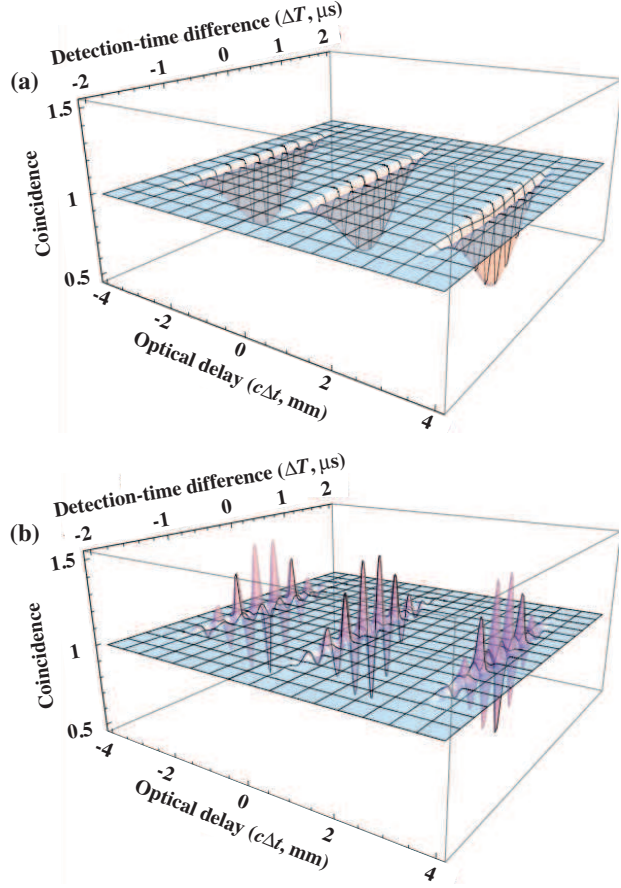}
\caption{Simulations of two-photon interference. Gaussian distributions of $|\gamma(\Delta t)|$ and $|\Gamma(\Delta T)|$ with the full width at half maximums of $0.67~{\rm ps}$ and $1.18~\micro\second$, respectively, are assumed. The periodic first-order coherence function for a diode laser with period of $T_p=10.57~{\rm ps}$ is also assumed. (a) For $\Delta\omega=0$, HOM dips with visibility of 0.5 can be shown when $\Delta T=0$ and  $\Delta t=n L_p$, where $n=0,\pm1,\pm2,\cdots$. (b) For $\Delta\omega=2\pi\times3~{\rm MHz}$, the coincidence shows the sinusoidal two-photon beating with respect to $\Delta T$ with envelopes defined by the two coherence functions.}
\label{theory}
\end{figure}

When $\Delta\omega\neq 0$, one can measure a sinusoidal oscillation with respect to $\Delta T$. It is notable that the coincidences always have the minimum at $\Delta T=0$ because Eq.~(\ref{I_3I_4_simple}) is independent of $\Delta\phi$. The visibility of the oscillation is determined by $|\gamma(\Delta t)|^2$. When $|\gamma(\Delta t)|=1$, one can measure the sinusoidal oscillation with $V=0.5$. As the frequency of the oscillation is determined by the difference between the frequencies of $E_1$ and $E_2$, $\Delta\omega$, the origin of the oscillation is a two-photon beating. As we qualitatively investigated earlier, the two-photon interference in terms of $\Delta T$ indeed reveals beating fringes with limited visibility.

From Eq.~(\ref{I_3I_4_simple}), one can expect an infinite sinusoidal oscillation with respect to $\Delta T$ as long as $|\gamma(\Delta t)|\neq 0$. In practice, we can consider a finite oscillation by considering a finite coherence time in terms of $\Delta T$. If we can somehow regulate the coherence between $E_1(t)$ and $E_1(t+\Delta T)$, we can consider the degree of first-order coherence between $E_1$ at time $t$ and $t+\Delta T$ as, $0\le|\Gamma(\Delta T)|\le1$, and it gives a finite interference in terms of $\Delta T$. Using $|\Gamma(\Delta T)|$, Eq.~(\ref{I_3I_4_simple}) can be modified as
\begin{equation}
\langle I_3I_4\rangle(\Delta t, \Delta T)\sim1-\frac{1}{2}|\gamma(\Delta t)|^2|\Gamma(\Delta T)|\cos[\Delta\omega \Delta T].
\label{I_3I_4_simple2}
\end{equation}
Note that we express Eq.~(\ref{I_3I_4_simple2}) as a function of $\Delta t$ and $\Delta T$ rather than $T_1$ and $T_2$ since it is dependent on these variables.

Equation~(\ref{I_3I_4_simple2}) is affected by $|\gamma(\Delta t)|$ and $|\Gamma(\Delta T)|$. An interesting model for the first-order coherence of the inputs $|\gamma(\Delta t)|$ is that it can be revived for a certain condition. It is known that multi-mode diode lasers show this property and the revival period is related to the finite cavity length of the diode laser. With the recurrence period of $T_p$, we can find the recurrence relation of the first-order coherence, $|\gamma(\Delta t)|=|\gamma(\Delta t+T_p)|$ where $t_c\ll T_p$ \cite{baek07}. Note that down-converted photons pumped by a multi-mode diode laser have a similar property \cite{kwon09}. With the recurrence property, let us assume that both the degrees of coherence, $|\gamma(\Delta t)|$ and $|\Gamma(\Delta T)|$, locally have a Gaussian shape with the conditions of $|\gamma(0)=|\Gamma(0)|=1$.

The numerical results of Eq.~(\ref{I_3I_4_simple2}) are visualized in Fig.~\ref{theory} (a) for $\Delta\omega=0$ and (b) for $\Delta\omega=2\pi\times 3~{\rm MHz}$. The full widths at half maximum (FWHM) of the Gaussian $|\gamma(\Delta t)|$ and $|\Gamma(\Delta T)|$ are assumed to be $0.67~{\rm ps}$ and $1.18~\micro\second$, respectively, and $T_p=10.57~{\rm ps}$. When $\Delta\omega=0$, we can see repeated HOM dips while varying $\Delta t$. The visibility of the HOM dips is a maximum at $\Delta T=0$, and as $\Delta T$ increases, the visibility decreases. For $\Delta\omega\neq0$, we can see sinusoidal oscillations rather than simple dips with respect to $\Delta T$. It is remarkable that we can attain either HOM peaks or dips while scanning of optical delay $\Delta t$ according to the detection time difference $\Delta T$: If $\Delta T$ is chosen so as to have minimum(maximum) coincidences, HOM dips(peaks) would appear in terms of $\Delta t$. Note that the repeated two-photon interference is expected when $\Delta t$ are multiples of $T_p$ for both the $\Delta\omega=0$ and $\Delta\omega\neq0$ cases.


\section{Experiment and result}

\begin{figure}[t]
\centering
\includegraphics[width=5.3in]{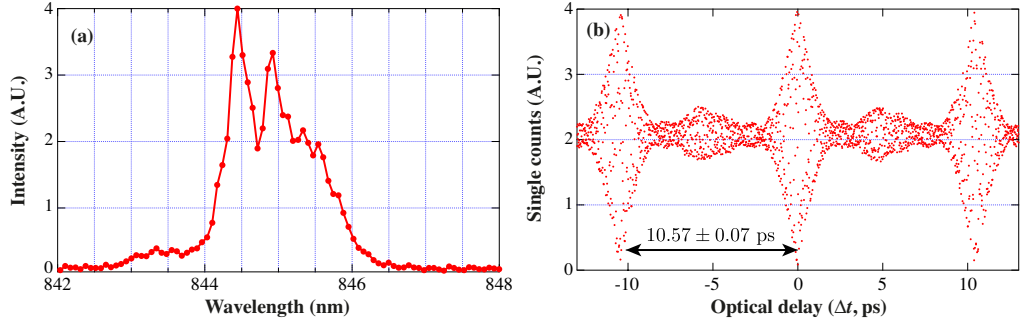}
\caption{(a) The spectrum of CW multi-mode diode laser. (b) Single-photon interference registered at D1 with synchronized AOMs.}
\label{MZ}
\end{figure}

In order to experimentally investigate the two-photon interference with CW coherent light, we built an experimental setup as shown in Fig.~\ref{setup} (b). A multi-mode diode laser at the wavelength of  845~nm was used. The FWHM of the wavelength is about 1~nm, see Fig.~\ref{MZ} (a). In order to maintain a non-zero $|\gamma(\Delta t)|$ while randomizing their phases, we built a Mach-Zehnder (MZ) interferometer with two BS and an optical delay. Two independent AOMs disturb the phase coherence between the two inputs while maintaining $|\gamma(\Delta t)|$. After the phases are randomized, Fig.~\ref{setup} (b) can be considered as Fig.~\ref{setup} (a) by considering the second BS of Fig.~\ref{setup} (b) as the BS of Fig.~\ref{setup} (a).

One can consider the interferometer as a regular MZ interferometer if AOM1 and AOM2 are synchronized since they conserve the phase coherence between the two arms of MZ interferometer \cite{kim13}. Under this condition, we measured the single-photon interference while scanning $\Delta t$, see Fig.~\ref{MZ} (b). It shows a clear recurrence of the MZ interference envelops with a period of $T_p=10.57\pm0.07~{\rm ps}$ \cite{baek07}.

When the driving radio frequency (RF) signals of the two AOMs are independent, the phase coherence between the two inputs would be disturbed. It is remarkable that unsynchronized RF signals can also introduce non-zero frequency difference between the two inputs, $\Delta\omega\neq0$, since an AOM adds the spectral frequency to the deflected beam according to the driving RF signal frequency. The RF signal frequency difference between AOM1 and AOM2, $\Delta f$, introduces $\Delta\omega=2\pi\Delta f$. Note that additional frequency modulation (FM) noise input to an AOM will disturb the phase within the same arm, thus degrading $|\Gamma(\Delta T)|$ as $\Delta T$ increases \cite{kim13}.

\begin{figure}[t]
\centering
\includegraphics[width=5.3in]{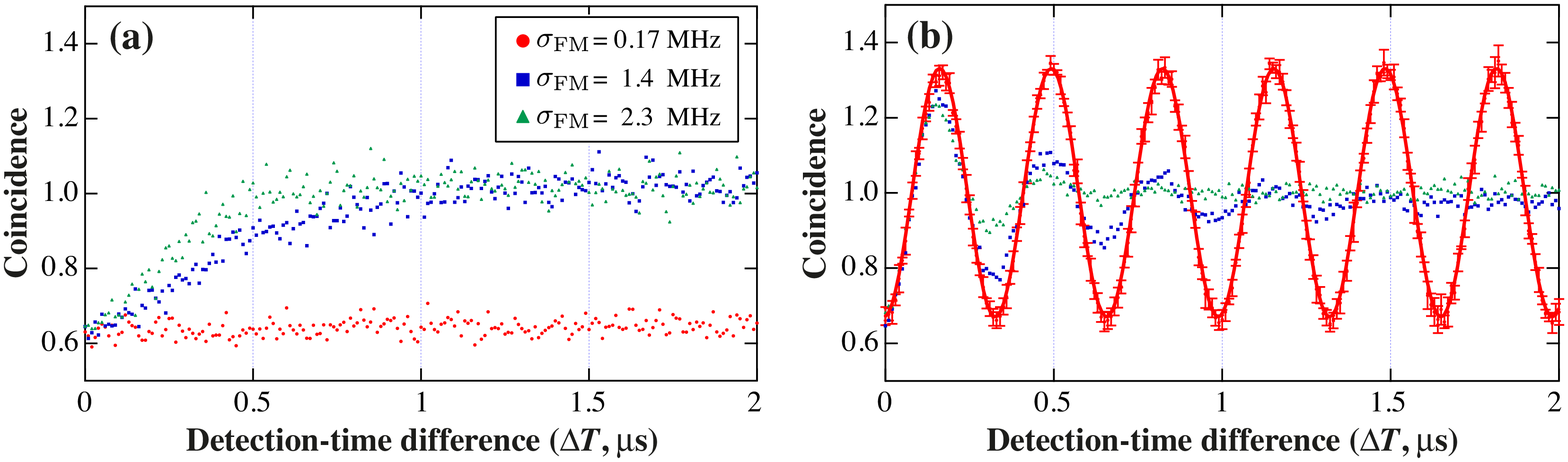}
\caption{Two-photon coincidences as a function of the detection-time difference $\Delta T$. The optical path length difference $\Delta t=0$ during the measurement. (a) $\Delta\omega=0$, (b) $\Delta\omega=2\pi\times3~{\rm MHz}$. (a) and (b) are measured at the various FM noises for RF signal of AOM1 that introduces limited $|\Gamma(\Delta T)|$. }
\label{histogram}
\end{figure}

While the single-photon interference is washed out, we measured the coincidences between D1 and D2 under various conditions. First, we measured the coincidence as a function of the detection time difference $\Delta T$ while $\Delta t$ is fixed at 0. Fig.~\ref{histogram} presents the coincidence for (a) $\Delta\omega=0$, and (b) $\Delta\omega=2\pi\times3~{\rm MHz}$. The TCSPC window is chosen to be 10~ns. We also present the data with different FM noise signals which degrade $|\Gamma(\Delta T)|$ when $\Delta T$ is large. The FM noise signals are quantified by the standard deviations of the RF signal frequency, $\sigma_{\rm FM}$. Note that we present the data with coincidence counts normalized so that the coincidence without two-photon interference is 1. Since the simulation result depicted in Fig.~\ref{theory} is also normalized in the same manner, this data presentation is convenient to compare theory and experiment.

Let us first discuss the result in Fig.~\ref{histogram} (a), in which $\Delta\omega=0$. In general, the coincidences show a regular HOM dip regardless of $\Delta T$. Note that each data point can be considered as a HOM dip since $\Delta t=0$, corresponding to the optical path length difference where a HOM interference occurs. The only exception to this general statement is that when $\sigma_{\rm FM}$ is sufficiently large, the visibility of the HOM dips decrease as $\Delta T$ increases and the visibility decreases faster when $\sigma_{\rm FM}$ is larger. From Fig.~\ref{MZ}, we can calculate the coherence time of our diode laser to be about $t_c=2.4~\pico\second$. Noting that the time scale in Fig.~\ref{histogram} is $\micro\second$, we find that two-photon interference occurs even when $t_c\ll\Delta T$. Interestingly, this  holds even if we input FM noise in order to degrade $|\Gamma(\Delta T)|$.

When $\Delta\omega\neq0$, the coincidences show a sinusoidal oscillation which corresponds to a two-photon beating fringe as depicted in Fig.~\ref{histogram} (b). Similar to the $\Delta\omega=0$ case, the oscillation continues even when $t_c\ll\Delta T$. The envelop of the oscillation is determined by the FM noise and they are identical to the $\Delta\omega=0$ case. The oscillation frequency is found to be 3.02~MHz which corresponds to the RF signal frequency difference.

\begin{figure*}[t]
\centering
\includegraphics[width=5.3in]{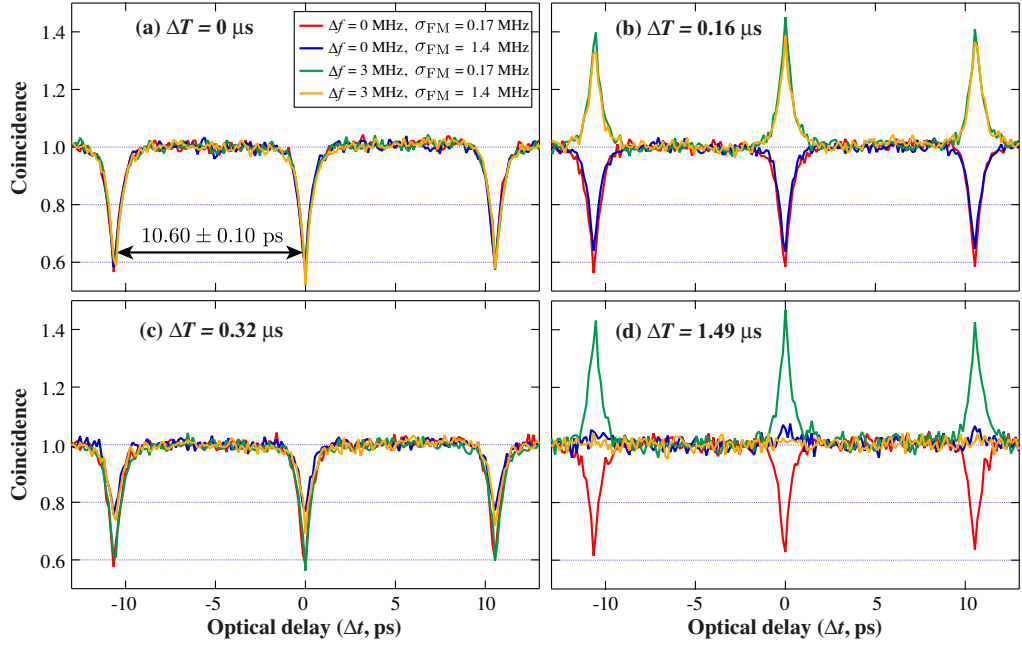}
\caption{Two-photon coincidences as a function of the optical path difference, $\Delta t$ for various detection-time differences (a) $\Delta T=0$, (b) $\Delta T=0.16~\micro\second$, (c) $\Delta T=0.32~\micro\second$, and (d) $\Delta T=1.49~\micro\second$. HOM dips/peaks are measured depending on the conditions. The dips/peaks are repeated every $T_p=10.60\pm0.10~ {\rm ps}$ which is the same for the single-photon interference repeating period.}
\label{hom}
\end{figure*}

After observing the two-photon interference in terms of $\Delta T$, we measured HOM interference, that is a two-photon interference as a function of the optical delay $\Delta t$ at a certain fixed detection-time difference $\Delta T$. We plot the normalized coincidences at four different $\Delta T$ in Fig.~\ref{hom}. We measured the coincidences under various $\Delta f$ which is related to $\Delta\omega$ and/or $\sigma_{\rm FM}$. Let us first consider the case when $\Delta T=0$ which is depicted in Fig.~\ref{hom} (a). Since the two photons which cause the coincidences at D1 and D2 are close to each other, it resembles a regular HOM interference. Regardless of the conditions, in this case, we can always observe the repeating HOM dips with $V\le0.5$. The recurrence period of the interference is $T_p=10.60\pm0.10~ {\rm ps}$, which is the same as the single-photon interference recurrence period shown in Fig.~\ref{MZ}. It is interesting to note that with down-converted photons pumped by a multi-mode diode laser, the two-photon interference revival does not occur although the single-photon interference revives \cite{kwon09}.

For $\Delta T\neq0$, we can see either HOM dips or peaks depending on the conditions. In particular, when $\Delta\omega=0$, we can always observe the HOM dips although the visibility of the interference decreases as $\sigma_{\rm FM}$ increases. On the other hand, we can measure either HOM dips or peaks for $\Delta\omega\neq0$. These phenomena actually come from the oscillation of two-photon interference in terms of $\Delta T$. If we choose $\Delta T$ to correspond to a maximum coincidence, e.g., $\Delta T=0.16~\micro\second$ or $1.49~\micro\second$, one can see the HOM peaks rather than dips while scanning $\Delta t$. For a $\Delta T$ which corresponds to a minimum coincidences, e.g., $\Delta T=0.32~\micro\second$, HOM dips appear. Note that the visibility of the HOM dips/peaks are affected by the FM noise, thus for a large $\sigma_{\rm FM}$, the HOM dips/peaks are suppressed. It is worth noting that the repeating property of HOM dips/peaks is preserved.


\section{Conclusion}

We report two-photon interference with CW multi-mode coherent light. Even though the two-photon interference with CW light is nonintuitive as the timing is not well defined, the non-zero first-order coherence function can provoke two-photon interference. We show that two-photon interference as a function of the detection time difference can reveal two-photon beating with the visibility $V\le0.5$. While varying the optical delay of the interferometer, HOM dips or peaks are observed depending on the chosen detection time difference. The HOM dips/peaks are repeated whenever the optical delay are multiples of the first-order coherence revival period of multi-mode coherent light. These results help to understand the nature of two-photon interference and also can be useful for quantum information science.


\section*{Acknowledgments}
The authors thank Y.-W. Cho, Y.-S. Ra, H.-T. Lim, and Y.-H. Kim for useful discussions. YSK acknowledges the support from the KIST Institutional Program (2E24013).

\end{document}